\documentclass[12pt,twoside]{article}
\usepackage{fleqn,espcrc1}

\usepackage{graphicx}
\usepackage[figuresright]{rotating}

\newcommand{\be}{\begin{equation}}
\newcommand{\ee}{\end{equation}}

\newcommand{\bea}{\begin{eqnarray}}
\newcommand{\eea}{\end{eqnarray}}

\newcommand{\La}{\ensuremath{\Lambda}\phantom{ }}
\newcommand{\Lcal}{\ensuremath{{\cal L}}}
\newcommand{\Psib}{\ensuremath{\overline{\Psi}}}
\newcommand{\p}{\ensuremath{\partial}}

\title{The structure of \La hypernuclei}

\author{Christoph M. Keil, Frank Hofmann and Horst Lenske\address{
        Institut f\"ur Theoretische Physik, Universit\"at Gie\ss en \\
        Heinrich-Buff-Ring 16, D-35392 Gie\ss en, Germany}}

\begin{document}

\maketitle

\begin{abstract}
The density dependent relativistic hadron field theory is extended to also describe strange systems.
It is seen that the application to hypernuclei works extremely well. Important spin orbit effects 
in \La hypernuclei are studied and their experimental relevance is pointed out.
\end{abstract}

\section{Introduction}
Since hypernuclei are unique laboratories for investigations of the interaction between hyperons and nucleons
and even among hyperons a unified description of baryonic in-medium interactions is
an important requirement. Unfortunately the application of Dirac-Brueckner theory
to medium and high mass nuclei is technically not feasible. A rather satisfactory 
approach providing a practicable way to use
microscopically deduced meson exchange interactions is the density dependent relativistic
hadron (DDRH) field theory \cite{Fuchs:1995as}. This theory has been successfully applied to isospin nuclei
and was thus considered to be an excellent basis for our hypernuclear studies.

\section{The density dependent relativistic hadron field theory}
The DDRH field theory is a relativistic lagrangian field 
theory of baryons interacting through the exchange of mesons:
\bea
\label{Lagrangian}
\Lcal &=& \Lcal_{B} + \Lcal_{M} + \Lcal_{int} \nonumber\\
\Lcal_{B} &=& \Psib_{F} \left[ i\gamma_\mu\p^\mu
                              - \hat{M} \right] \Psi_{F} \nonumber\\
\Lcal_{M} &=&\frac{1}{2} \sum_{i = \sigma, \sigma_s}
\left(\p_\mu\Phi_i\p^\mu\Phi_i - m_{i}^2\Phi_i^2\right)
           - \frac{1}{2} \sum_{\kappa = \omega, \phi, \rho, \gamma}
             \left( \frac{1}{2} F^{(\kappa)^2} - m_\kappa^2 A^{(\kappa)^2}
\right) \label{eq:ModLagr} \\
\Lcal_{int} &=& \Psib_F \hat{\Gamma}_\sigma(\Psib_F, \Psi_F) \Psi_F \sigma
- \Psib_F \hat{\Gamma}_\omega(\Psib_F, \Psi_F) \gamma_\mu \Psi_F \omega^\mu
- \frac{1}{2}\Psib_F \hat{\vec\Gamma}_\rho(\Psib_F, \Psi_F) \gamma_\mu \Psi_F
\vec\rho^\mu \nonumber\\
&&+ \Psib_F \hat{\Gamma}_{\sigma_s}(\Psib_F, \Psi_F) \Psi_F \sigma_s
- \Psib_F \hat{\Gamma}_\phi(\Psib_F, \Psi_F) \gamma_\mu \Psi_F \phi^\mu
- e \Psib_F \hat{Q} \gamma_\mu \Psi_F A^\mu, \nonumber
\eea
where $\Psi_F$ denotes a spinor consisting of the octet baryons. To include the medium effects 
the coupling constants are
treated as functionals depending on lorentz scalar bilinear products of the baryonic field 
operators, ensuring lorentz invariance and thermodynamical consistency \cite{Fuchs:1995as}.
Different to conventional RMF theory \cite{Serot:1986ey} self-interactions are introduced
on the level of baryonic correlations.

\subsection{Density dependent interactions}
To obtain the vertex functionals Dirac-Brueckner-Hartree-Fock (DBHF) self-energies are calculated 
in infinite nuclear 
matter using realistic interactions, in our case the Bonn A NN-potential \cite{Brockmann:1990cn}. 
These are decomposed in the different meson channels and then
mapped onto Hartree-Fock mean field self-energies:
$
\Gamma_\alpha\left(\hat\rho\right) \Phi_\alpha = \Sigma^{DB}_\alpha.
$
This enables us to perform a nuclear structure calculation with standard methods but applying a
microscopically deduced interaction. A detailed description of the DDRH procedure can be found in 
\cite{Hofmann:2000vz,Keil:2000hk}.

\subsection{Medium effects}
There are two minor and one major change in the equations of motion of the fields
due to the functional structure of the vertices:
\begin{itemize}
\item The source terms in the mesonic equations of motion: 
$ \left( \partial^2+m^2 \right)\Phi = \Gamma_\Phi \left( \hat\rho \right) \rho_\phi$
acquire an additional medium dependence in the coupling.
\item The baryonic self-energies get an additional density dependence (DD), but also get an extra term which
does not appear in usual RMF:
$
\frac{\delta \mathcal L_{int}}{\delta \overline{\psi}} = 
\Sigma^{(0)}\left(\hat\rho\right)\psi +\frac{\partial \mathcal L_{int}}{\partial \hat\rho}
\frac{\delta \hat\rho}{\delta \overline{\psi}}.
$
This is the so called {\it rearrangement self-energy} and accounts for static polarizations of the
nuclear medium \cite{Fuchs:1995as}.
\end{itemize}

\subsection{Extension into the strange sector}
The extension to the strange sector would be rather straightforward if there were DBHF calculations
for the whole baryonic octett available, but, unfortunately, they are not. However, an analysis of the
DB equations reveals that, assuming the exchange of only net flavor neutral mesons as in the nuclear
case \cite{Keil:2000hk}:
\begin{enumerate}
\item The functional shape of the $\Gamma_{B\alpha}\left(\hat\rho\right)$ is approximately equal
for a given $\alpha$ and the relative scaling factor is approximately given by the ratio of the
free space couplings:
\bea
&\Sigma_{B\alpha}\left( \rho \right) = \Gamma_{B\alpha}\left( \rho \right)\Phi_\alpha 
\Rightarrow \Gamma_{Y\alpha}\left( \rho \right) =\Gamma_{N\alpha}\left( \rho \right)
\frac{\Sigma_{Y\alpha}\left( \rho \right)}{\Sigma_{N\alpha}\left( \rho \right)}& \\
&R_{Y\alpha} \equiv  \frac{\Sigma_{Y\alpha}\left( \rho \right)}{\Sigma_{N\alpha}\left( \rho \right)} =
\frac{g_{Y\alpha}}{g_{N\alpha}}\left( 1+\mathcal O\left( 1-\frac{M_N}{M_Y} \right) \right) 
\approx \mbox{const.}& \nonumber
\eea
\item The DD is approximately determined by the respective baryon density.
\end{enumerate}
This enables us to set up a microscopic model for \La hypernuclei.

\subsection{A model for single \protect\La hypernuclei}
\begin{figure}[htb]
\begin{minipage}[t]{80mm}
\includegraphics[width=78mm]{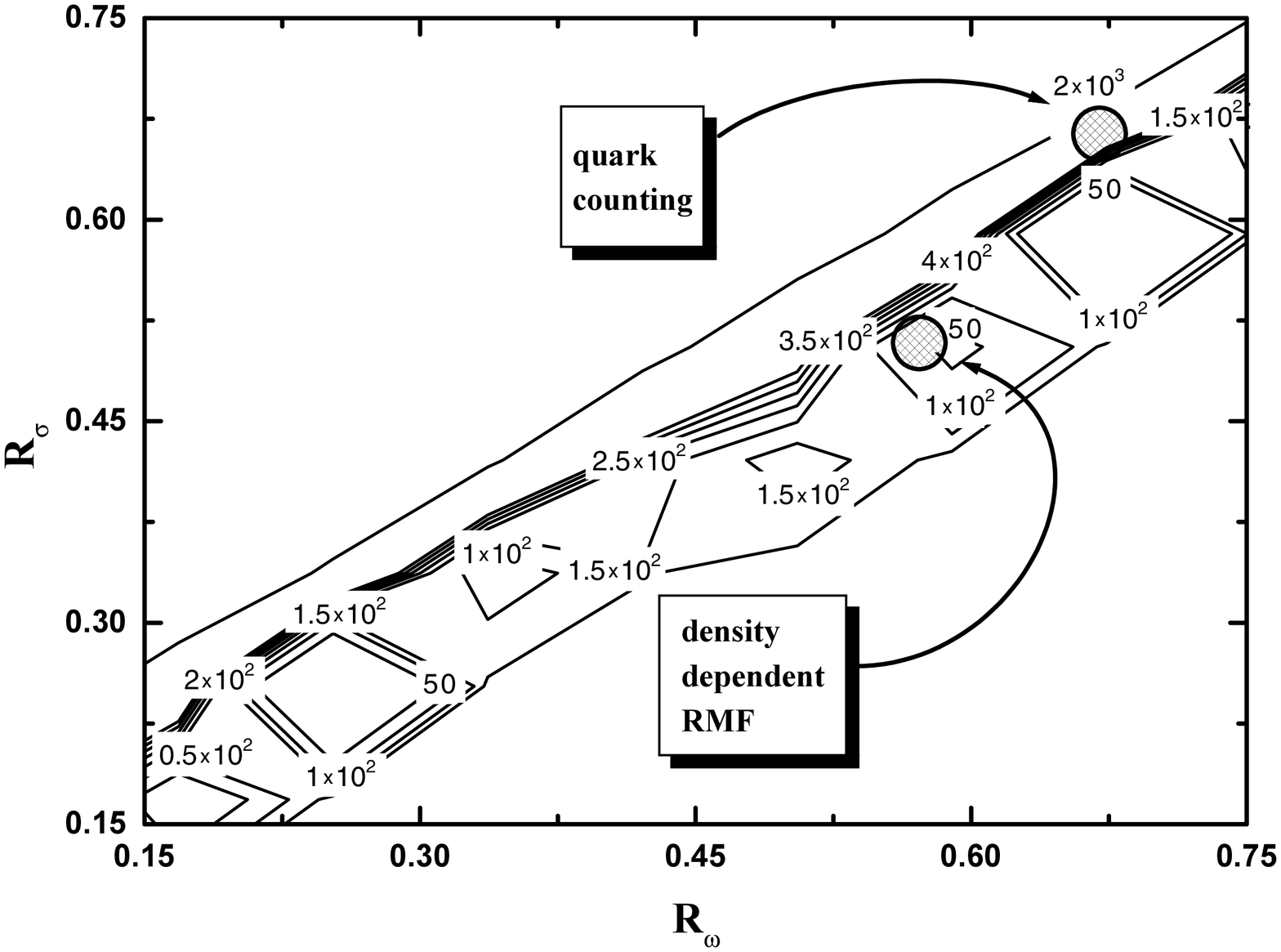}
\caption{$\chi^2$ analysis of theoretical vs. experimental single particle energies \protect\cite{Keil:2000hk}.}
\label{fig:chi2}
\end{minipage}
\hspace{\fill}
\begin{minipage}[t]{75mm}
\includegraphics[width=75mm]{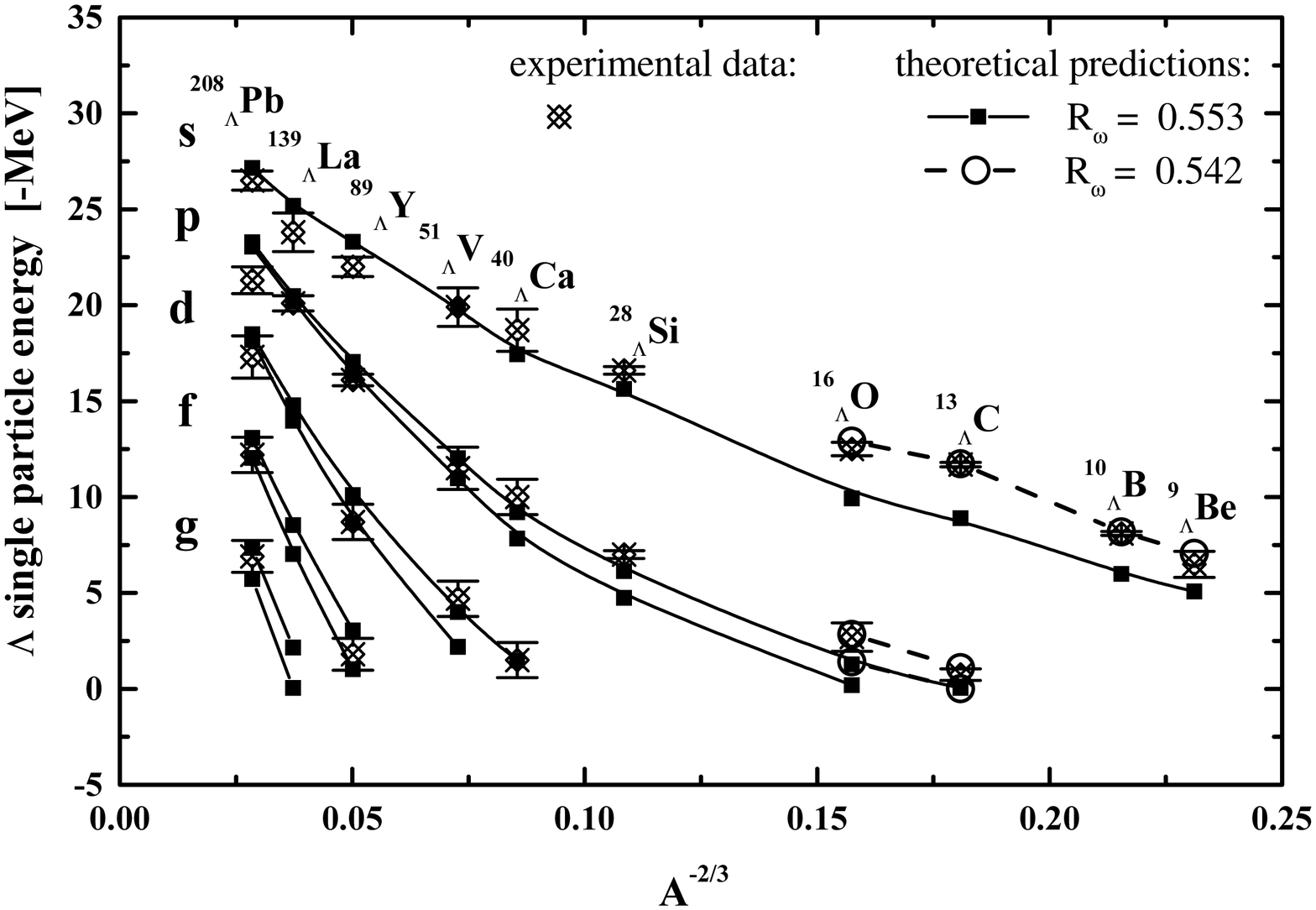}
\caption{Comparison of experimentally measured and calculated single particle spectra}
\label{fig:spec}
\end{minipage}

\end{figure}

In the baryonic sector the model contains neutrons, protons and the $\Lambda$, whereas the mesonic sector
consists of the $\sigma$-, the $\omega$- and the $\rho$-meson. The isoscalar mesons are treated in
the above described DD fashion. A model for the \La is especially attractive since it
only couples to the isoscalar mesons for which the relative couplings R$_{\Lambda\sigma}$
and R$_{\Lambda\omega}$ have to be determined now. A least squares fit of the scaling factors with respect
to experimental \La single particle spectra does not yield an unique choice but only a sharp, deep 
``valley'' of equally well choices, see figure~\ref{fig:chi2}~\cite{Keil:2000hk}. The quark counting choice of 
$\frac{2}{3}$ for both relative couplings is clearly ruled out. To be as microscopic as possible we
chose R$_{\Lambda\sigma}$ according to a T-matrix calculation by Haidenbauer et al. 
\cite{Haidenbauer:1998kk} and determined R$_{\Lambda\omega}$ via our $\chi^2$ analysis. Experimantal
data for medium to heavy hypernuclei with well resolved spin orbit (s.o.) splitting would be an essential
constraint to  narrow the correlation range of acceptable \{R$_{\Lambda\sigma}$,R$_{\Lambda\omega}$\}
pairs

\section{Results}
\subsection{The \protect\La mean field}
The \La central potential in our model has a depth of roughly -30 MeV what is in good agreement with experimental
analyses. The shape is comparable to the neutron potential leading to fairly delocalized \La wave
functions (i.e. in $^{40}_{\Lambda}$Ca the rms radii of the d shell are well above that of the 
whole nucleus, 3.6 fm compared to 3.3 fm). This leads to a reduction of the s.o. splitting, as 
discussed in section~\ref{sec:partprop}.

Due to the DD of the interaction the central potential of the lambdas is modified compared 
to that of standard RMF moels. The DD couplings yield an effective radial dependence of the
coupling strength, i.e. a reduction in the nuclear center with respect to the surface region. A second
effect is due to the rearrangement self-energies which add a repulsive part to the central potential. Since
the \La density in single \La hypernuclei is fairly low these two effects are not very large. The dominant
contribution of the DD description is due to the surrounding nuclear environment which is also
described in DDRH. Here the static polarization effects of the nuclear medium play an essential role. The
fine structure of the \La single particle spectra is a unique signature of the surrounding nucleonic 
environment.

\subsection{Single particle properties}
\label{sec:partprop}
\begin{figure}[htb]
\begin{minipage}[t]{75mm}
\includegraphics[width=75mm]{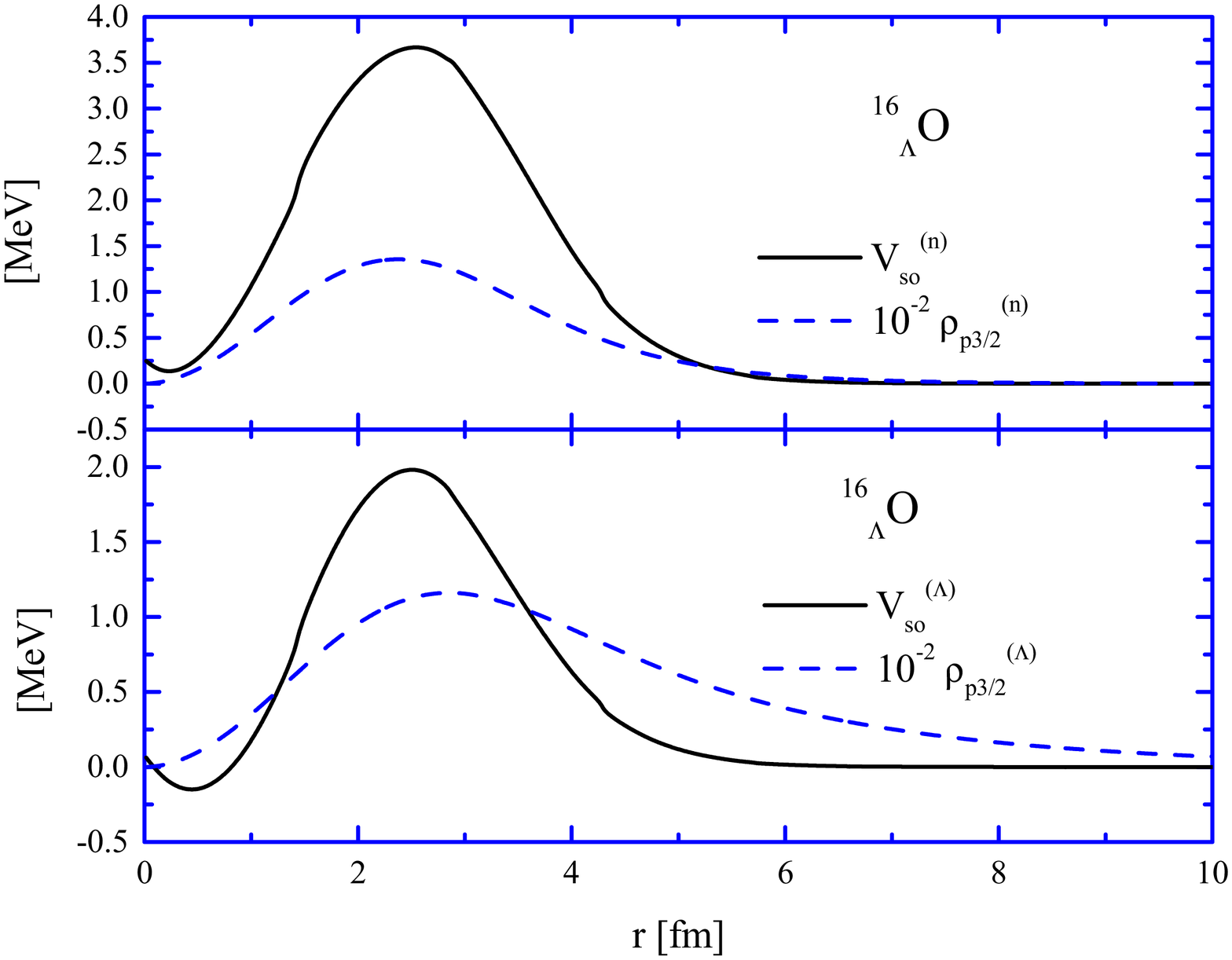}
\caption{Overlap of the single particle density with the s.o. potential.}
\label{fig:so-overlap}
\end{minipage}
\hspace{\fill}
\begin{minipage}[t]{80mm}
\includegraphics[width=70mm]{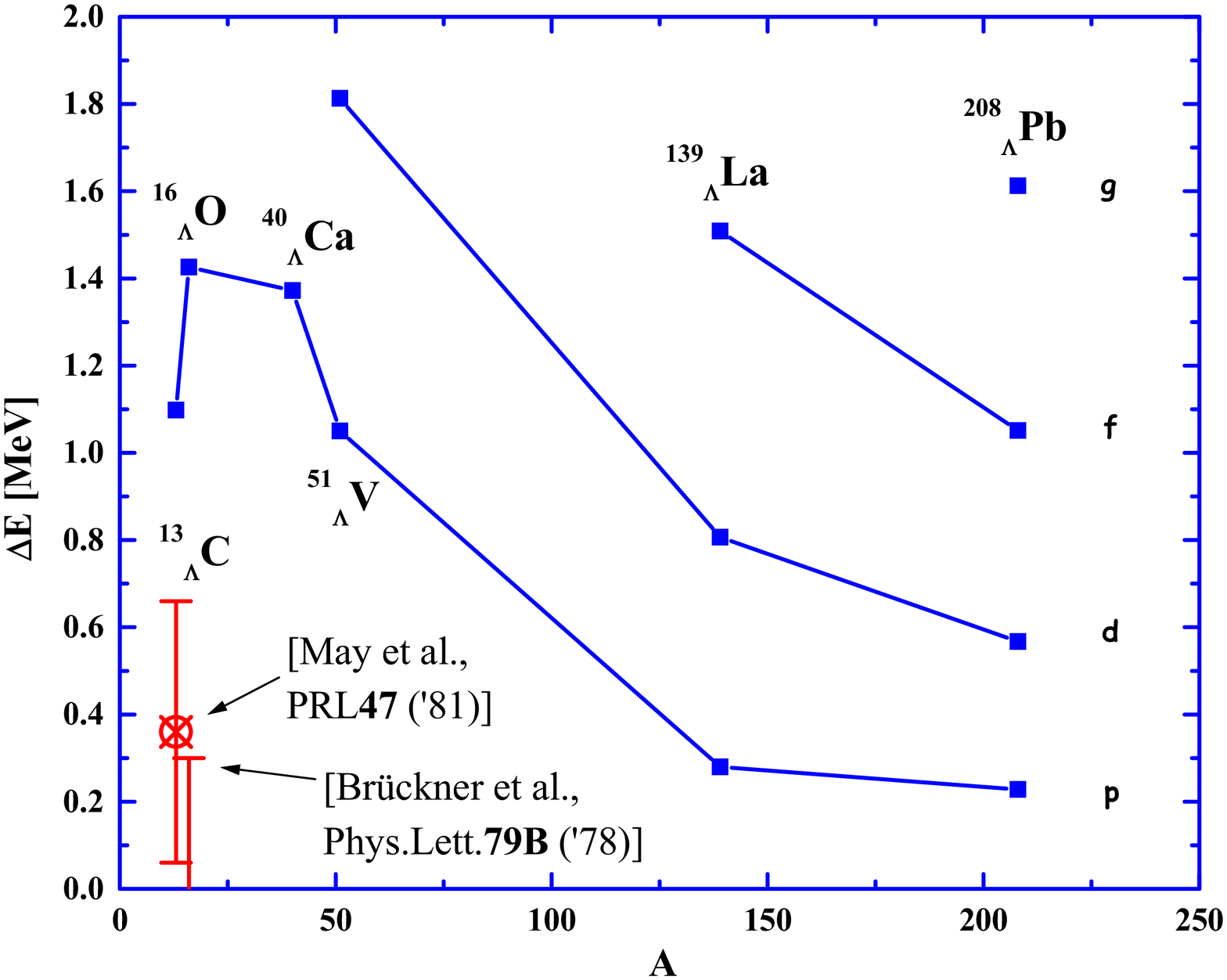}
\caption{A dependence of the s.o splitting. Native s.o. effects are diluted for A$<$40 due to continuum
threshold effects.}
\label{fig:so-split}
\end{minipage}
\end{figure}
As is shown in figure~\ref{fig:spec} the description of large to medium heavy hypernuclei is excellent.
For the surface dominated light nuclei a slight correction of R$_\omega$ has to be made to adjust for
e.g. the presence of clusters in light nuclei which the model does not contain.

In general the \La single particle spectra resemble in many aspects the neutron spectrum which is shifted in
energy and has a squeezed level structure. But in addition to that the \La s.o. splitting shows a
reduction due to the largely delocalized wave functions of the $\Lambda$. This effect is shown in 
figure~\ref{fig:so-overlap}. The delocalization of the \La reduces the overlap of the wave function with
the s.o. potential, which is relatively sharply localized on the nuclear surface, and thereby the
s.o. energy splitting is reduced stronger than expected from scaling. Another prominent s.o. effect 
is shown in figure~\ref{fig:so-split}.
Besides the decrease of the s.o. splitting when going to large mass numbers (since s.o. splitting is 
a finite size effect) there is also a decrease for nuclei below $^{40}_\Lambda$Ca. This
is due to a squeezing of the single particle energy spectrum for states close to the continuum 
threshold. Unfortunately, all the high resolution experiments have been made for light mass hypernuclei
and thus do not reveal too much native spin orbit information.

\section{Conclusions and outlook}
With upcoming high precision experimental data the microscopic properties of the extended DDRH can be
used to extract important information of NN and YN interactions. The model has also been successfully 
applied to neutron star matter including hyperons \cite{Hofmann:2000mc}.

\end{document}